\documentclass[iop]{emulateapj}

\usepackage[english]{babel}
\usepackage{amsmath,amsfonts,amssymb}
\shorttitle{On the Geometry of the IBEX Ribbon}
\shortauthors{Sylla and Fichtner}

\begin{document}

%% LaTeX will automatically break titles if they run longer than
%% one line. However, you may use \\ to force a line break if
%% you desire.

\title{On the Geometry of the IBEX Ribbon }

%% Use \author, \affil, and the \and command to format
%% author and affiliation information.
%% Note that \email has replaced the old \authoremail command
%% from AASTeX v4.0. You can use \email to mark an email address
%% anywhere in the paper, not just in the front matter.
%% As in the title, use \\ to force line breaks.

\author{Adama Sylla, Horst Fichtner}
\affil{Institut f\"ur Theoretische Physik IV, 
       Ruhr-Universit\"at Bochum, 44780 Bochum, Germany}
%Universit\"atstra\ss{}e, 44780 Bochum, Germany} 

%% Notice that each of these authors has alternate affiliations, which
%% are identified by the \altaffilmark after each name.  Specify alternate
%% affiliation information with \altaffiltext, with one command per each
%% affiliation.

%\altaffiltext{1}{Visiting Astronomer, Cerro Tololo Inter-American Observatory}
%CTIO is operated by AURA, Inc.\ under contract to the National Science
%Foundation.}
%\altaffiltext{2}{Society of Fellows, Harvard University.}
%\altaffiltext{3}{present address: Center for Astrophysics,
%    60 Garden Street, Cambridge, MA 02138}
%\altaffiltext{4}{Visiting Programmer, Space Telescope Science Institute}
%\altaffiltext{5}{Patron, Alonso's Bar and Grill}

%% Mark off your abstract in the ``abstract'' environment. In the manuscript
%% style, abstract will output a Received/Accepted line after the
%% title and affiliation information. No date will appear since the author
%% does not have this information. The dates will be filled in by the
%% editorial office after submission.

\begin{abstract}

The Energetic Neutral Atom (ENA) full-sky maps obtained with the Interstellar Boundary Explorer (IBEX) show an unexpected bright narrow band of increased intensity. This so-called ENA ribbon results from charge exchange of interstellar neutral atoms with protons in the outer heliosphere or beyond. Amongst other hypotheses it has been argued that this ribbon may be related to a neutral density enhancement, or H-wave, in the local interstellar medium. Here we quantitatively demonstrate, on the basis of an analytical model of the  principal large-scale heliospheric structure, that this scenario for the ribbon formation leads to results
that are fully consistent with the observed location of the ribbon in the full-sky maps at all energies  detected with high-energy sensor IBEX-Hi.
\\ 
\end{abstract}

\section{Introduction}
The Interstellar Boundary Explorer (IBEX) has provided the first energy-resolved all sky maps of the flux of energetic neutral
atoms (ENAs). These IBEX maps reveal, above a general solar wind-structured ENA flux background, a `ribbon' of increased flux. 
Several models have been proposed to explain the source, location, and structure of the ribbon.

\citet{McComas-etal-2009b} and \citet{Schwadron-etal-2009} were the first to suggest that the ribbon might result from 
consecutive charge-exchange processes. This scenario has, subsequently, been modelled quantitatively \citep[see, e.g.,][]{Heerikhuisen-etal-2010, 
Heerikhuisen-Pogorelov-2011, Strumik-etal-2011, Moebius-etal-2013, Schwadron-McComas-2013, Zirnstein-etal-2013, Burlaga-etal-2014, Heerikhuisen-etal-2014, 
Isenberg-2014}.
Despite the model's basic success to explain the ENA ribbon, there remain critical open questions regarding the stability of the (pick-up ion) 
seed distribution of the ribbon ENAs in the local interstellar medium \citep{Florinski-etal-2010, Gamayunov-etal-2010, Burlaga-Ness-2014}.
Other suggestions comprise the ideas that the source regions of the ribbon ENAs are located far beyond the heliopause (HP) at the edge of
the local interstellar cloud \citep{Grzedzielski-etal-2010} or rather inside the heliosphere \citep{Fahr-etal-2011b, Kucharek-etal-2013,
Siewert-etal-2013}. A different scenario that involves the magnetic and neutral density structure of the local interstellar medium (LISM) on 
the one hand, but assumes the production region of the ENAs to be mainly in the inner heliosheath (IHS) on the other hand was suggested recently
by \citet{Fichtner-etal-2014}.

All of these hypotheses have been summarized and critically assessed in detail in the review papers by
\citet{McComas-etal-2014a} and \citet{McComas-etal-2014b} with the result that, while it is clear that the ENAs establishing the IBEX ribbon are
related to the interaction of the heliosphere with the local interstellar medium, there is no consensus yet on their source region(s).

In this paper we follow up on the idea that the IBEX ENA ribbon is a result of a so-called H-wave (section 2) transiting through the heliosphere 
\citep{Fichtner-etal-2014}. In order to translate that idea into a quantitative model, we construct the geometry of the ribbon (section 3) within
the framework of a simple but well-suited model of the principal large-scale heliospheric structure. We discuss a best fit to the ENA data obtained with the 
IBEX-Hi detector along with the significance for the findings regarding the relation between the ribbon and the local interstellar magnetic 
field (section 4) and the sensitivity of the results to parameter changes (section 5). Section 6 contains a brief summary of results and the 
conclusions regarding the H-wave hypothesis. 

\section{Model of the heliosheath and the H-wave induced ribbon formation}
\begin{figure}[!b]
\begin{center}
 \includegraphics[width=0.47\textwidth]{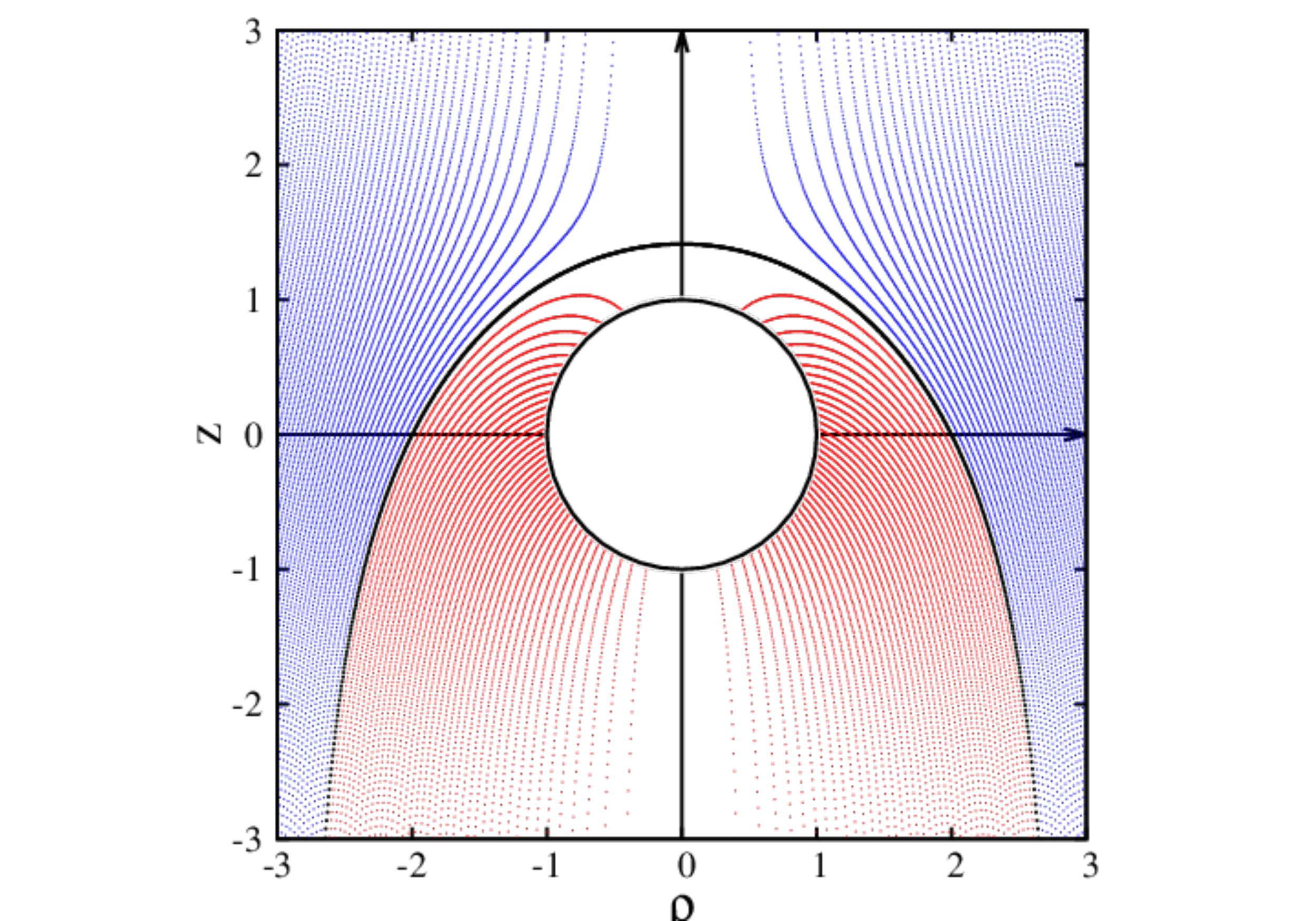}
 \caption{The flow lines in the inner and outer heliosheath (see Eq.\ 4). The outer thick black line is the heliopause ($\eta=1$, see Eq.\ 5).
          The red color corresponds to inner heliosheath ($\eta<1$) and the blue color to outer heliosheath ($\eta>1$). The inner circle
          is the termination shock.}
\end{center}
\end{figure}
\subsection{Plasma flow}
For the present purpose both the solar wind plasma flow in the IHS and the interstellar plasma flow in the outer heliosheath (OHS), 
i.e.\ in the regions between the termination shock (TS) and the heliopause (HP) and outside the latter (formally up to the interstellar
bow shock), can be described as being incompressible \citep[${\nabla\cdot\bf{u}}=0$, see][]{Roeken-etal-2015}:  
\begin{eqnarray}
  {\bf  u} & = &  - {\nabla  \Phi} 
             = - \left(\frac{\partial \Phi}{\partial \rho}\right)\, {\bf {e}}_{{\boldsymbol\rho}} 
               - \left(\frac{\partial \Phi}{\partial x_3 }\right)\, {\bf e}_{ \bf x_3}\\
           & = & \frac{ k \rho u_{LISM}}{ r^3}\,{\bf e}_{\boldsymbol\rho}-\frac{u_{LISM}}{r^3} (r^3 - k x_3)\,{\bf e}_{\bf x_3}
\end{eqnarray}
with the scalar velocity potential in cylindrical coordinates 
\begin{eqnarray}
{\Phi (\rho, x_3)} = u_{LISM} x_3 + \frac{k u_{LISM}}{r}
\;\;\;;\;\;\; k=const. 
\end{eqnarray}
where $r = \sqrt{{\rho}^2+ {x^2_3}}$, $\rho = \sqrt{{x^2_1}+{x^2_2}}$, and $(x_1,x_2,x_3)$ denote Cartesian coordinates and $u_{LISM}$
is the speed of the undisturbed LISM flow. In this formulation $k u_{LISM}$ is interpreted as the speed of the shocked solar wind in 
the IHS. The resulting flow lines $x_3(\rho)$ can be obtained from the equation 
\begin{eqnarray*}
\frac{dx_3}{d\rho}=\frac{u_{x_3}}{u_ {\rho}} 
\end{eqnarray*}
or from the associated stream function (see Appendix A) with the solution:
\begin{equation} 
  {x_3(\rho)} = \left(\eta -\frac{{ \rho}^2}{2 k} \right)  \rho \left(1-\left[ \eta -\frac{{ \rho}^2}{2 k} \right]^2 \right)^{-\frac{1}{2}}
  \end{equation}
The parameter $\eta$ is identifying the flow lines and $k$ characterizes the relative strength of the solar and the interstellar wind, which 
we take as $k=2$ \citep[see][]{Roeken-etal-2015}. With this and $\eta=1$ we obtain the following formula for the HP surface 
\begin{eqnarray} 
 x_3 (\rho) = \left(1 -\frac{{ \rho}^2}{4} \right) \rho \left(1-\left[ 1 -\frac{{\rho}^2}{4} \right]^2 \right)^{-\frac{1}{2}} 
\end{eqnarray}
Figure~1 illustrates the resulting flow lines in the IHS and the OHS. The interstellar flow comes from the positive $z$-direction. 
The black lines indicate the HP $\eta=1$ (see Eq.\ 5) and the TS, respectively. For simplicity, we assume the latter to
be a Sun-centered sphere. The red lines in the IHS correspond to $\eta<1$, the blue ones in the OHS to $\eta>1$ (see Eq.\ 4). 
\subsection{Structure in the neutral gas}
\citet{Fichtner-etal-2014} have summarized the arguments supporting the view that the LISM is inhomogeneous and is likely to exhibit propagating 
wave- or pulse-like structures. These authors argued that the waves in the plasma must be expected to induce (via charge exchange coupling)
associated structures in the neutral gas \citep[see also][]{Shaikh-Zank-2010}.
In particular, as a consequence of a slow wave in the plasma (that propagates along the magnetic field 
that is oriented as sketched in the top panel of Fig.~2, see \citet{McComas-etal-2009}), there should be a wave in the neutral gas, too, a so-called H-wave. 
\begin{figure}[t]
 \includegraphics[width=0.47\textwidth]{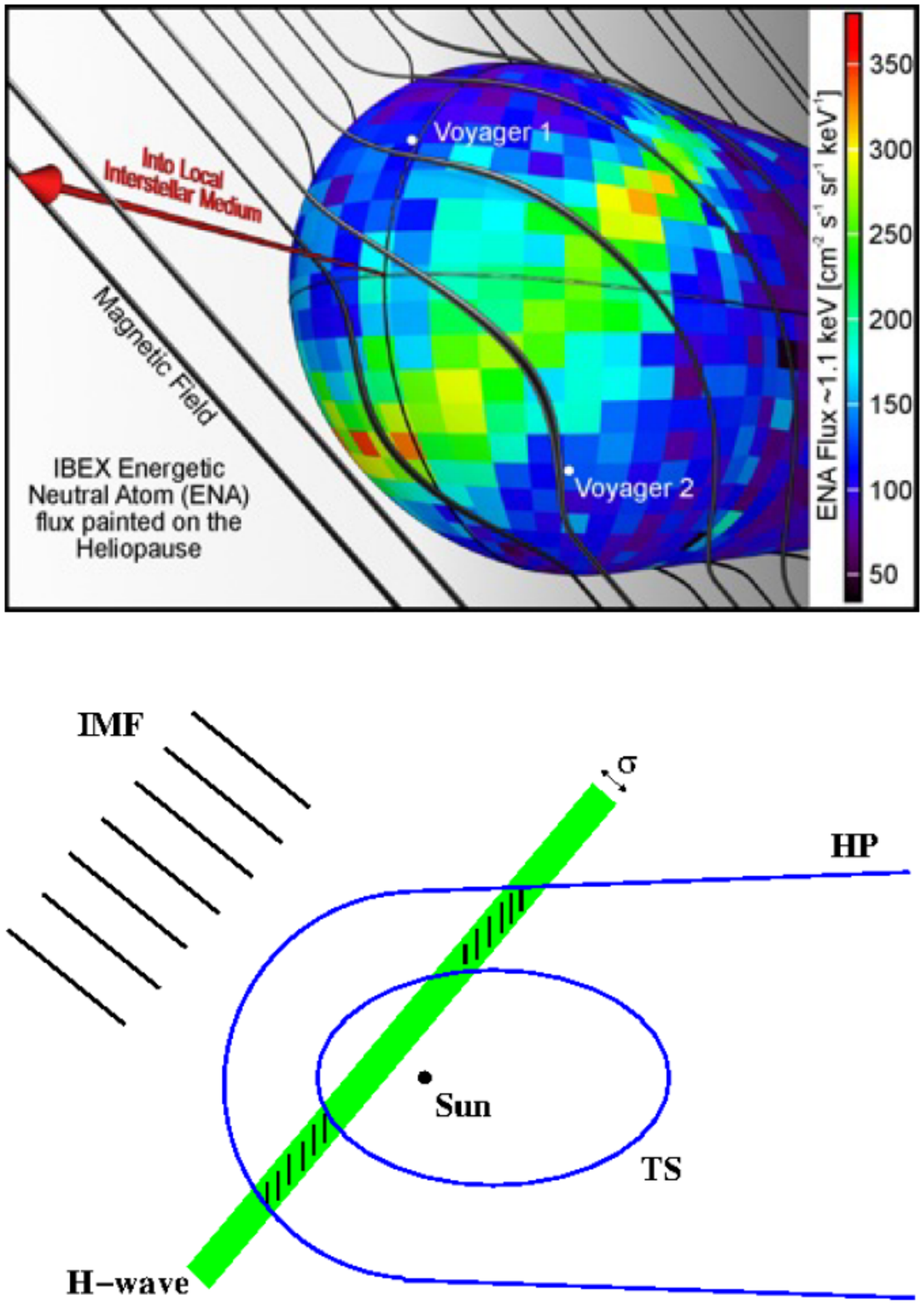}
 \vspace*{0.8cm}\\
\caption{Top panel: The flux of energetic neutral atoms projected onto the heliopause around which the interstellar magnetic field lines are 
 draping (adopted from McComas et al. 2009b).
 Bottom panel: Sketch of the H-wave scenario in a plane perpendicular to the orientation of the undisturbed interstellar magnetic field 
 (IMF, black lines). An enhancement of interstellar density, i.e.\ an H-wave of thickness $\sigma$ is propagating through the heliosphere that is depicted by the
 terminaton shock (TS) and the heliopause (HP). In the intersection region, which is indicated by the shaded areas and which forms a ring-like structure in 3D, the production rate of ENAs is increased due to the higher neutral density.}
\end{figure} 
The bottom panel of Fig.~2 shows a sketch of such an H-wave, representing an enhancement of interstellar gas density that propagates along 
the interstellar magnetic field and, after decoupling from the plasma that flows around the `obstacle' heliosphere, penetrates the heliosphere.
While in this sketch the H-wave front is perpendicular to the undisturbed interstellar magnetic field, not only this but also other orientations
are quantitatively studied below.  
The production rate of ENAs is highest in the IHS \citep[see, e.g.][]{Sternal-etal-2008} and directly proportional to the neutral density
\citep[e.g.][]{Fahr-etal-2007} so that one must expect an increased ENA flux to be generated in the shaded regions shown 
in the bottom panel of Fig.~2. 
\section{Description of the ribbon geometry} 
\begin{figure}[!t]
\begin{center}
\includegraphics[width=0.47\textwidth]{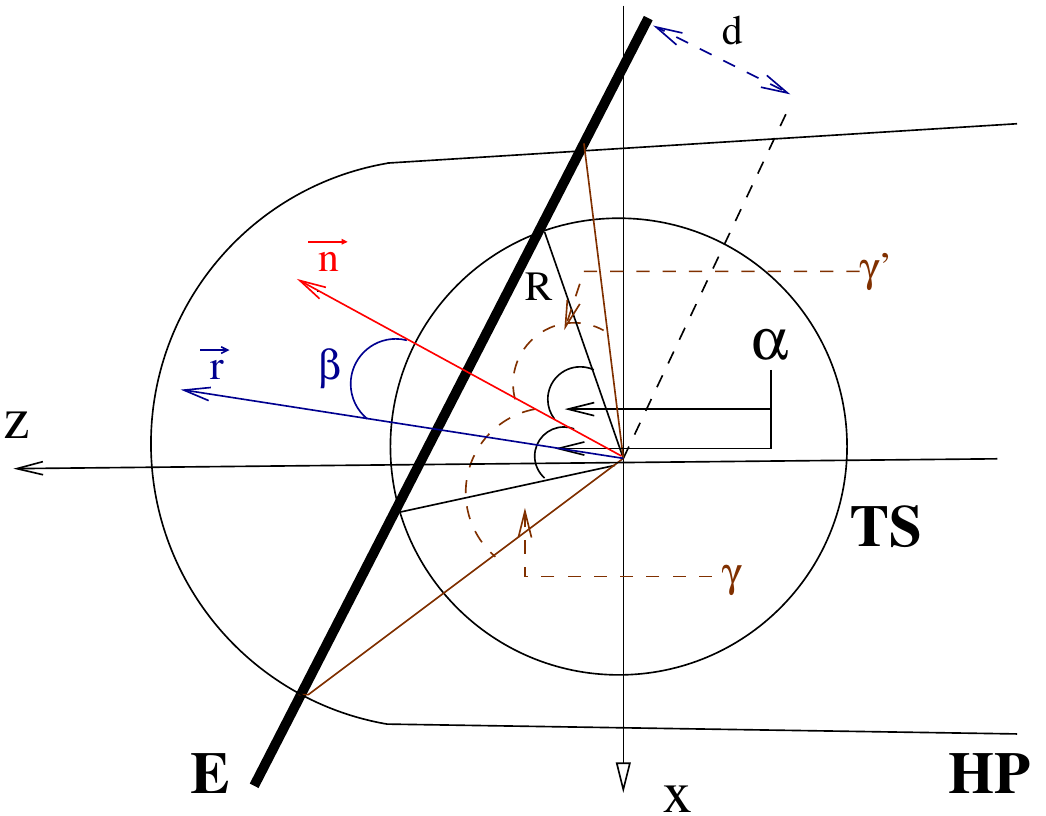}
\end{center}
\caption{Geometry of the intersection of the H-wave with the heliosphere: In the chosen plane the termination shock (TS) is the inner circle, the
         heliopause (HP) is the outer curve and the H-wave is indicated by the plane E with a normal vector ${\bf n}$. 
         The angles $\alpha$, $\gamma$, and $\gamma^\prime$ represent respectively the 'inner boundary' and 'outer boundary' of the ribbon and, thus, 
         its angular width. For more details see text.}
\end{figure}
Figure~3 illustrates the basic geometry of the ribbon in the scenario suggested by \citet{Fichtner-etal-2014}. The TS is assumed as a Sun-centered 
sphere with radius $R$, the HP as an axisymmetric surface defined by Eq.\ (5), and the H-wave is indicated as the thick black line E. 
The angle between the directions to the intersection of E with the TS and the normal vector $\bf n$ is denoted by $\alpha$, defining 
the 'inner boundary' of the ribbon in a given plane:
\begin{equation}
\alpha = \arccos \left( \frac {d}{R} \right)
\end{equation}
~\\
Here $d$ denotes the shortest heliocentric distance to the plane E. The 'outer boundary' of the ribbon is defined by the directions to the intersections
of the plane E with the HP. In a given plane those directions can be specified in terms of the angles $\gamma$ and $\gamma^\prime$. If ${\bf r}_{HP}$ and
${\bf r}^{\prime}_{HP}$ denote the heliocentric directions to these intersections the corresponding angles follow from 
\begin{equation}
\gamma^{(\prime)}= \arccos \left[ \frac{{\bf {r}^{(\prime)}}_{HP} \cdot \bf {n}} { |{\bf {r}^{(\prime)}}_{HP} ||\bf {n}|} \right]
\end{equation} 

Given the chosen symmetrical TS, the angle $\alpha$ is the same for both 'sides' of the ribbon in a given plane, while the angles ${\gamma}$ and
${\gamma'}$ are different. With $\alpha$, $\gamma$ and $\gamma^{\prime}$ it is now straightforward to formulate the condition that a given line of sight
intersects the ribbon: 
\begin{eqnarray} 
\alpha \leq \beta \leq \gamma,\gamma^{\prime}
\end{eqnarray} 
where $\beta$ is the angle between the chosen heliocentric direction ${\bf r}$ and the normal ${\bf n}$ (see Fig.~3), i.e.
\begin{eqnarray}
   \beta &=& \arccos \left( \frac {\bf{r} \cdot \bf{n}}{|\bf{r}||\bf{n}| } \right)\nonumber\\
         &=& \arccos \left( \frac {x_1 n_1+x_2 n_2+x_3 n_3}{\sqrt{(x^2_1+x^2_2+x^2_3)(n^2_1+n^2_2+n^2_3)} } \right)
\end{eqnarray}
where in the Cartesian coordinates as defined in section~2.1:
\begin{eqnarray} 
  {\bf{n}}=({n_1}, {n_2}, {n_3})  \;\;\;  \hbox{\rm and}   \;\;\;    {\bf{r}}=({x_1}, {x_2}, {x_3}) 
\end{eqnarray}
With these formulas we are now in the position to calculate the location and angular width of the ribbon in the all-sky ENA flux maps as observed 
with the IBEX spacecraft.
\begin{figure}[!t]
\begin{center}
~\vspace*{0.0cm}\\
\includegraphics[width=0.50\textwidth]{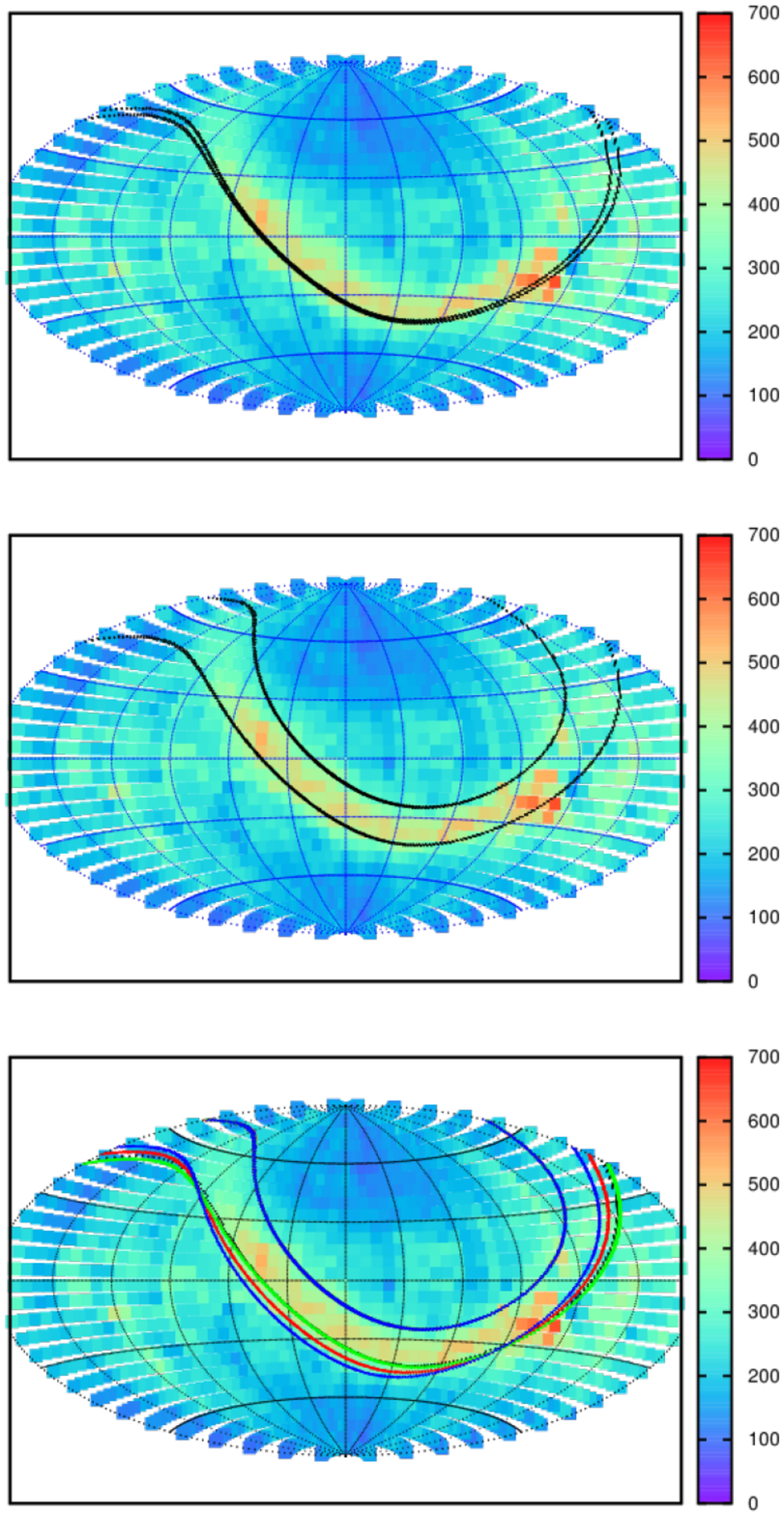}
\vspace*{-0.5cm}\\
\caption{\footnotesize The observed ENA ribbon in an all-sky map of the observed ENAs fluxes (ENAs/(cm$^2$~s~sr~keV) at 0.71~keV vs.\ different model results:
         Top panel: The ribbon (two thick black lines) as resulting from an infinitely thin H-wave with $\sigma = 0$, i.e.\ a plane E
         with $d = 0.1 R$ (see Fig.~3). Note that also for this case the ribbon has a finite angular width because the angles $\alpha, \gamma$, and
         $\gamma^{\prime}$ are different.
         Middle panel: The ribbon resulting from an H-wave with finite thickness $\sigma = d_2-d_1= 0.4R-0.1R=0.3R$ is the region between the black lines 
         indicating its inner (upper) and outer (lower) boundary. 
         Bottom Panel: The ribbon location and thickness resulting from a transformation of the heliopause defined with Eq.(5): While the latter results in 
         a lower boundary indicated by the green line, one obtains the red and blue lines when it is tilted by 5 and 10 degrees, respectively. The untilted but 
         polynomial heliopause (see text) leads to almost identical result (black line) than Eq.(5), i.e.\ the green line.
         For these and all following all-sky maps, the data of which is available via the IBEX data release website 
         http://ibex.swri.edu/researchers/publicdata.shtml, an Aitoff-projection was used such that the heliospheric nose direction is in the center of the map
         (see Appendix B).}
\end{center}
\end{figure}
\section{The resulting ribbon geometry}
For the visualisation of the resulting band of higher ENA fluxes in the all-sky maps we use an Aitoff projection (see Appendix B) for all following
figures. As a first step we check on the principal location and width of the ribbon originating from an H-wave, i.e.\ we choose 
its thickness $\sigma = d_2-d_1$ (with $d_{1,2}$ denoting the shortest heliocentric distances of its sunward and anti-sunward side, 
respectively) and its orientation given by the normal vector ${\bf n}$.

The top panel of the Fig.~4 gives the result of an infinitely thin H-wave (i.e.\ with vanishing thickness $\sigma$) oriented perpendicularly to the
vector ${\bf n} = (-1.8,1.3,1.5)$, which is 
corresponding to the unit vector ${\bf n}/\vert{\bf n}\vert \approx (-0.672,0.485,0.560)$ and which 
is anti-parallel to the most likely direction of the undisturbed local interstellar magnetic field (see the discussion in section~5 below).
Its shortest heliocentric distance is $d = 0.1 R$. Note that although the H-wave is infinitely thin, the corresponding ribbon is not: In an all-sky map 
it has a finite angular width because the angles $\alpha, \gamma$, and $\gamma^{\prime}$ are different, as is evident from the sketch in Fig.~3.  
The middle panel shows the result for an H-wave with the same orientation but a finite width
(as sketched in Fig.~2), namely $\sigma = d_2 - d_1 = 0.4 R - 0.1 R = 0.3 R$. The inner (upper in the all-sky map) angular ribbon boundary is determined by the 
angle $\alpha$ resulting from the intersection of the anti-sunward plane with the TS, and the outer (lower) boundary by the angles $\gamma, \gamma^{\prime}$
resulting from the intersection of the sunward plane with the HP, as described in section~3. 
The bottom panel shows the effect of different transformations of the HP on the lower ribbon boundary.
First, a transformation to a polynomial shape is achieved by a Taylor expansion of the HP function defined with Eq.(5),
for details see Appendix C. Comparing the green line (resulting from Eq.(5)) and the black line (Taylor-expanded HP function) reveals that the effect of a 
polynomial HP on the ribbon location and width is negligible.
This can easily be understood as a consequence of the fact that, in the upwind heliosphere
($z \geq 0$), even in lowest (second) order the Taylor expansion is well approximating the HP defined with Eq.(5), see Figure~7 in appendix~C. Second, a tilt 
of the HP by 5 or 10 degrees, achieved by subsequent rotations about the $x$- and $y$-axis, results in the lower boundary indicated by the red and blue line, 
respectively. These tilt angles are motivated by the recent finding by \citet{Wood-etal-2014} that the heliotail direction is deviating at most by 10 degrees 
from the inflow direction of the LISM. Evidently, the effect is significant and one concludes (i) from a comparison of the red and the green line that a tilt
of the HP improves the agreement between the modelled and the observed ribbon and (ii) from a comparison of the red and the blue line that the tilt must be 
expected to be less than 10 degree, consistent with the findings by \citet{Wood-etal-2014}. Therefore, for all following computations, we used the HP function 
Eq.(5) with an additional tilt by 5~degrees.  

The results shown in Fig.~4 make it evident that the H-wave hypothesis results in an ENA ribbon at the correct location in the all-sky maps and that its width
must be a few tens of AU, as was already speculated in \citet{Fichtner-etal-2014}. The given geometrical H-wave parameters were iterated such that the resulting
ribbon geometry is a simultaneous best fit to the high-energy ENA maps at 1.11, 1.74, and 2.73~keV provided by the IBEX-Hi sensor. 
The result for these are shown in Fig.~5, which demonstrates that the H-wave induced ENA ribbon geometry is generally consistent with the IBEX measurements at
all these energies. Note that for the highest energy channel at 4.3~keV shown in the bottom panel the ribbon feature is clearly present and well-fitted at high
northern latitudes (top of the map), but that there are significant ENA emissions outside the ribbon at lower latitudes. These additional signals at higher energies
are also known from measurements with the INCA instrument aboard Cassini \citep{Krimigis-etal-2009}. An explanation of the associated `broadening' of the ribbon at
higher energies probably requires invoking additional ENA sources, like secondary ENAs as, e.g., discussed in \citet{Heerikhuisen-etal-2014}.

\begin{figure}[!t]
\begin{center}
\includegraphics[width=0.50\textwidth]{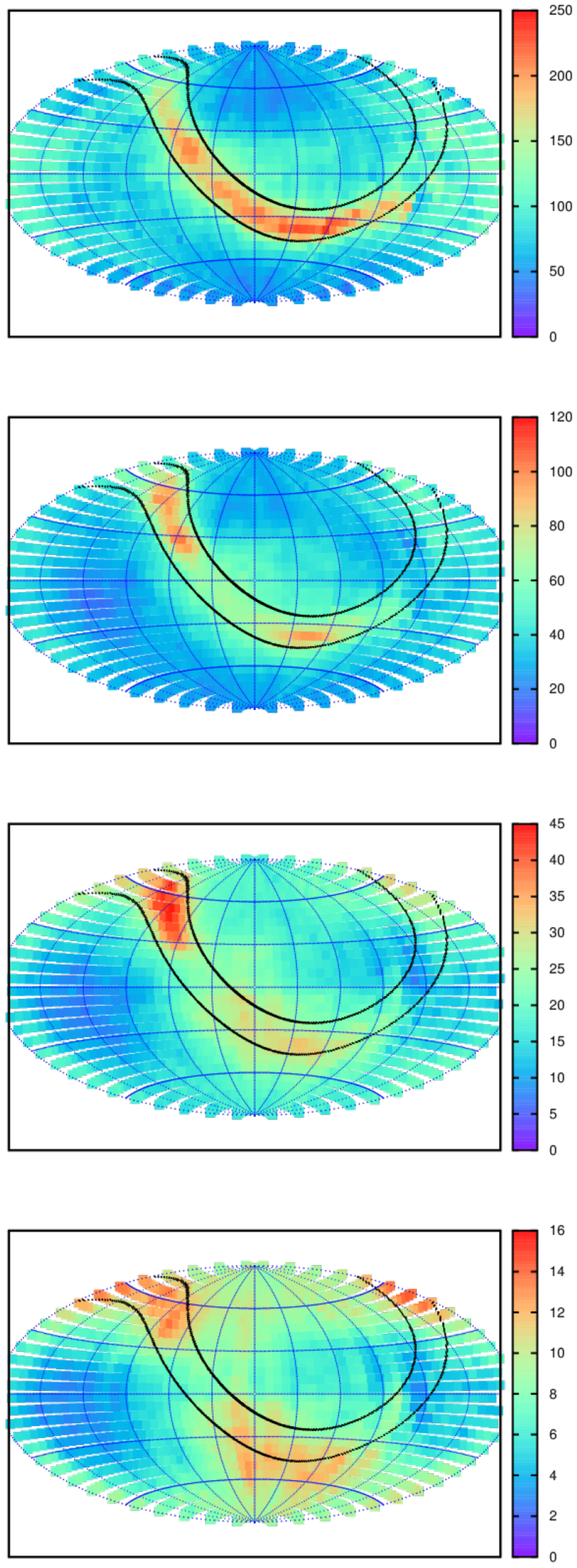}
\vspace*{-1.0cm}\\
\end{center}
\caption{All-sky maps of the ENA fluxes (ENAs/(cm$^2$~s~sr~keV) as observed by IBEX-Hi and the simulated best-fit
         (see text) ribbon geometry for the energies 1.11 keV, 1.74 keV, 2.73 keV, and 4.29 keV from top to bottom.} 
\end{figure}

The above findings corroborate the assumption that the ENAs forming the ribbon in the IBEX-Hi all-sky maps
can indeed originate in the intersection region of an H-wave with the inner heliosheath. While this 
represents a rather different explanation from all other scenarios that have been suggested, so far,
it does confirm the relation of the ribbon to the local interstellar magnetic field that is needed 
in most other scenarios, too. One must 
distinguish, however, in the present model between the field direction and the normal vector to 
the H-wave front: As discussed in \citet{Fichtner-etal-2014} it is, in principle, possible that the
H-wave front is not perpendicular to the magnetic field. The effect of a different orientation along
with different widths of the neutral density enhancement is discussed in the next section. 

\section{Sensitivity to the H-wave parameters}
\begin{figure*}[!t]
\begin{center}
\includegraphics[width=0.95\textwidth]{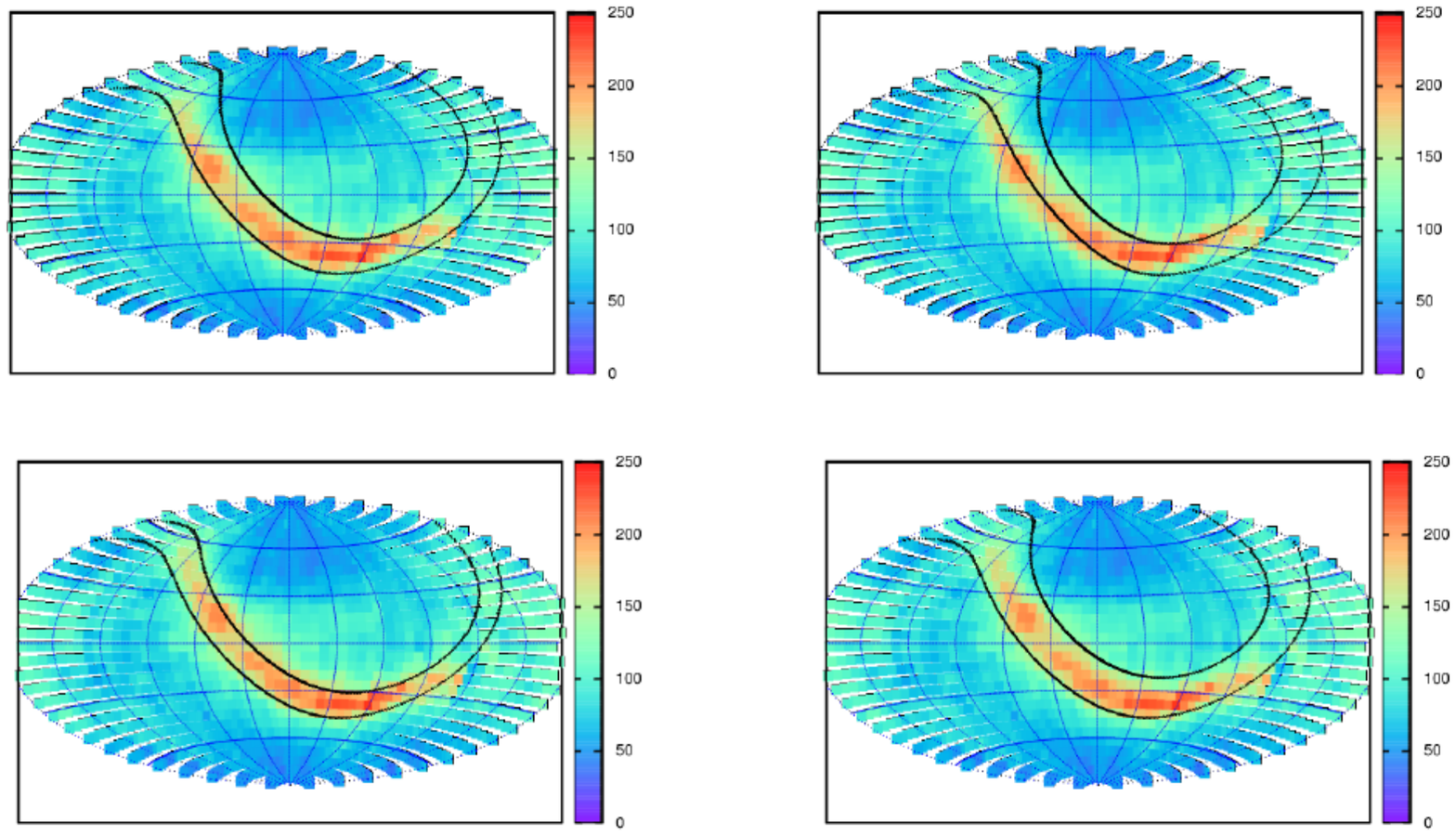}
\vspace*{-6.0cm}\\
\end{center}
\caption{All-sky maps of observed ENA fluxes (ENAs/(cm$^2$~s~sr~keV) at 1.1~keV and overlayed simulated ribbon geometry. The upper two plots show 
the result if the normal vector of the H-wave front is changed by 5 and 10 degrees from the best fit direction. 
The lower two plots display the results for the best fit direction but with the two different H-wave widths 
$\sigma = d_2-d_1 = 0.3 R - 0.1 R = 0.2 R$ and $\sigma = d_2 - d_1 = 0.45 R - 0.1 R = 0.35 $.} 
%~\vspace{0.6cm}\\
\end{figure*}
To illustrate the sensitivity of the result to the orientation of the H-wave front, i.e.\ to check on the goodness 
of the best fit, the upper two panels of Fig.~6 give the ribbon geometry for a wave normal vector ${\bf n}$ whose 
direction differs by 5 and 10 degree, respectively, from the best fit direction. While the first result (upper left panel)
is still compatible with the observational data, the second (upper right panel) is clearly not.

It interesting to note
that the best fit normal vector and the upwind direction $-{\bf u}_{LISM}$ have an angle of $(55\pm5)^o$, which is close to
the $49^o$ between the upwind direction for the untilted HP and the interstellar magnetic field estimated by \citet{Heerikhuisen-etal-2014}. 
The ecliptic longitude $\lambda_{ecl}\approx 205^o$ and latitude $\delta_{ecl}\approx 35^o$ of the best fit normal
vector are also slightly different than the values discussed in the literature \citep[e.g.,][]{Witte-etal-1996, 
Heerikhuisen-Pogorelov-2011, Borovikov-Pogorelov-2014, Wood-etal-2015}. Keeping in mind, however, the simplifying assumptions regarding
the TS and HP surfaces made in our analytical approach, one can safely state that the normal vector is closely related 
to the direction of the undisturbed local interstellar magnetic field. 

The two lower plots in Fig.~6 show the influence of the H-wave width and reveal that the region of increased neutral 
density as well as, in turn, of enhanced ENA production and flux must indeed be assumed to be about 25-28~AU wide, assuming a 
TS radius of 84 to 94~AU \citep{Stone-etal-2005, Stone-etal-2008}.

\section{Conclusion} 
In this paper we have constructed an analytic model that reproduces the correct geometry of the IBEX ribbon
in the all-sky ENA flux maps and, thereby, corroborates the hypothesis that a propagating localized density increase in
the neutral interstellar gas, termed an H-wave, can be the cause of the IBEX ribbon. The best fit of this geometry to 
IBEX ENA data depends particularly on the orientation and width of the H-wave whose transit through the heliosphere
leads to an increased production of ENAs in the inner heliosheath. Despite the simplifying assumptions regarding the 
termination shock and heliopause surfaces, the proposed scenario, although rather different from all others that 
have been invoked to explain the IBEX ribbon, makes it likely that the ribbon is closely related to the direction of
the undisturbed local interstellar magnetic field.

In subsequent work we will extend the modelling to a computation of the actual fluxes of ENAs resulting from an H-wave
intersecting the inner heliosheath, i.e.\ we will evaluate the relevant line-of-sight-integrals \citep[see, e.g.,][]{Sternal-etal-2008,
Fichtner-etal-2014} from an inner boundary (IB) at the detector to an outer boundary (OB) sufficiently beyond the ENA source 
region (i.e.\ in the present case beyond the HP):
\begin{eqnarray}  
\Phi\left(E_{\rm ENA},\vartheta,\varphi\right) 
  = \frac{1}{4\pi}\int\limits_{\mbox{\tiny IB}}^{\mbox{\tiny OB}}
      \left[n_{p}f_{p}\left(v_{p}\right) n_{H} \sigma_{ex} v_{rel}\right]
      ds
\label{ena-flux}
\end{eqnarray}  
with the solar wind and pick-up ion proton velocity ($v_p$) distribution function $f_{p}$ and number density $n_p$, the charge exchange 
cross section $\sigma_{ex}$ and the relative speed $v_{rel}$ between a proton and an interstellar neutral hydrogen atom.
There are two key ingredients: first, the proton velocity distribution function whose evolution has to be computed from a transport model
like in \citet{Fahr-Fichtner-2011} or \citet{Fahr-etal-2014}, but here for the IHS. The structure of $f_{p}$ in the intersection region 
of the H-wave and the IHS determines the `fine structure' of the ENA fluxes in the ribbon in the all-sky-maps at different energies.
Second, the ENA flux is directly proportional to the number density $n_H$ of the hydrogen atoms and, thus, to the H-wave signature. From
astronomical observations \citep[e.g.][]{Haverkorn-Goss-2007, Welty-2007} and corresponding simulations \citep{Hennebelle-Audit-2007} it 
is derived that $n_H$ can easily vary by a factor of two and more down to the few~AU scale in the so-called warm neutral (interstellar) 
medium. Since the latter reflects the properties of the neutral component in the LISM \citep[e.g.,][]{Stanimirovic-2009}, one can expect 
to see such variation as a local H-wave with a two- to threefold enhanced $n_H$. This directly translates, via Eq.(\ref{ena-flux}), into
a correspondingly increased ENA flux, which in turn represents the general ribbon feature in the all-sky maps.

\acknowledgments
We are grateful for discussions with Andrzej Czechowski, Frederic Effenberger, Jacob Heerikhuisen, 
Dave McComas, Eberhard M\"obius, Klaus Scherer, Nathan Schwadron, Gary Zank, and Ming Zhang. We 
also appreciate discussions at the team meeting `Heliosheath Processes and
Structure of the Heliopause: Modeling Energetic Particles, Cosmic Rays, and
Magnetic Fields' supported by the International Space Science Institute
(ISSI) in Bern, Switzerland.
%\bibliographystyle{apj}
%\bibliography{references}

\appendix

\section{A: Analytical calculation of the flow lines}
In cylindrical coordinates $(\rho=\sqrt{x_1^2+x_2^2}, x_3)$ the velocity potential (Eq.\ 3) reads
\begin{equation}
{ \Phi (\rho, x_3)} = u_{LISM} x_3 + \frac{ k {u_{LISM}}}{\sqrt {\rho^2+x^2_3}}  
\end{equation}
with $u_{LISM}$ denoting the velocity of the undisturbed local interstellar medium (LISM) and $k$ the relative
strength of the (shocked) solar and interstellar wind. The flow lines are on surfaces of constant stream function
$\Psi$. The latter can be derived from the incompressibility condition ${\nabla} \cdot {\bf u}=0$, which reads in
cylindrical coordinates explicitly:
\begin{equation}
0= { {\nabla} \cdot {\bf u}} = \frac{1}{\rho} \frac{\partial}{\partial \rho}\left(\rho u_{\rho}\right)
                             + \frac{\partial}{\partial x_3}\left(u_{x_3}\right) 
\end{equation}
From this it follows that the stream function must fulfil the two equations:
\begin{equation}
\frac{1}{\rho} \frac{\partial \Psi}{\partial x_3}= u_{\rho} 
\;\;\;\;; \;\;\;\;
\frac{1}{\rho} \frac{\partial \Psi}{\partial \rho}= u_{x_3}
 \end{equation}
which have the solutions:
\begin{equation}
\Psi = -\int {\frac{k u_{LISM} \rho^2}{\sqrt[3]{\rho^2+x^2_3}} dx_3 + G(\rho) }  
     = -\frac{k u_{LISM} x_3}{\sqrt{\rho^2+x^2_3}} + G(\rho)  
\end{equation}
and
\begin{equation}
\Psi = -\int{u_{LISM} \rho d\rho} + \int{ \frac{k u_{LISM} \rho^2}{\sqrt[3]{\rho^2+x^2_3}} d \rho+ F(x_3) } 
     = -\frac{1}{2}{u_{LISM}\rho^2} {-\frac{k u_{LISM}\rho^2}{\sqrt{\rho^2+x^2_3}}} + F(x_3)  
\end{equation}
with two functions $G(\rho)$ and $F(x_3)$ occuring as integration constants regarding integration w.r.t.\ 
$x_3$ and $\rho$, respectively. Chosing $G(\rho)=\frac{1}{2}u_{LISM}\rho^2$ and $F(x_3)=0$ leads to
\begin{equation}
\Psi(\rho,x_3) = {-\frac{k u_{LISM} x_3}{\sqrt{\rho^2+x^2_3}}} - \frac{1}{2}{u_{LISM}\rho^2}
\end{equation}
The condition $\Psi(\rho,x_3) = \tilde{\eta} = const$ describes the flow lines. With the definition 
$\eta=-\tilde{\eta}/(ku_{LISM})$ one has
\begin{equation}
\eta=  \frac{\rho^2}{2k}+\frac{x_3}{\sqrt{\rho^2+x^2_3}} 
\end{equation}
from which one finds the desired equation (4) for the flow lines
\begin{equation}
x_3(\rho) = \left(\eta -\frac{{ \rho}^2}{2 k}\right) \rho \left(1-\left[ \eta -\frac{{ \rho}^2}{2 k} \right]^2 \right)^{-\frac{1}{2}}
\end{equation}
\section{B: Coordinate transformation and Aitoff Projection}
The Cartesian coordinates introduced in section~2 are related via
\begin{equation}
x_1 = R\sin{\vartheta} \cos{\varphi}
\;\;\;;\;\;\; 
x_2 = R\sin{\vartheta} \sin{\varphi} 
\;\;\;;\;\;\; 
x_3 = R\cos{\vartheta}   
\end{equation}
to a heliocentric spherical polar coordinate system. In order to plot the desired 
all-sky maps centered on the heliospheric nose the following transformation is applied
\begin{equation}
x_1' =  x_3 = R\sin{\vartheta'} \cos{\varphi'}
\;\;\;;\;\;\; 
x_2' =  x_2 = R\sin{\vartheta'} \sin{\varphi'} 
\;\;\;;\;\;\; 
x_3' = -x_1 = R\cos{\vartheta'}   
\end{equation}
with the new latitude and longitude angles $\vartheta'$ and $\varphi'$. 

The Aitoff projection maps these spherical polar coordinates on Cartesian ones $(x,y)$ in a plane via
\begin{equation}
x = \frac{2 \alpha \cos(\phi) \sin\left(\frac\lambda 2\right)}{\sin(\alpha)} 
\;\;\;;\;\;\; 
y = \frac{\alpha\sin(\phi)}{\sin(\alpha)}
\;\;\;;\;\;\; 
\alpha = \arccos\left(\cos(\phi)\cos\left(\frac\lambda 2\right)\right)
\end{equation}
The dependence of the angles $\phi$ and $\lambda$ on $\vartheta'$ and $\varphi'$ as well as on $\vartheta$ and $\varphi$ is given by
\begin{equation}
\phi=\frac{\pi}{2} -\vartheta' = \arcsin(-\cos{\varphi}\sin{\vartheta}) 
\;\;\;;\;\;\; 
\lambda = \varphi' = \arccos(\cos{\vartheta}/\cos{\phi}) 
\end{equation}
and they must be interpreted as the latitude and longitude from the central meridian, respectively. 
\section{C: Polynomial heliopause function}
The Taylor expansion of the HP function Eq.(5) up to sixth order reads:
\begin{equation}
x_3(\rho) = \sqrt{2} - \frac{3\sqrt{2}}{16}\,\rho^2 - \frac{5\sqrt{2}}{512}\,\rho^4 - \frac{7\sqrt{2}}{8192}\,\rho^6 + O(\rho^8)
\end{equation}
The corresponding `polynomial' HP surfaces are illustrated with Fig.~7 where the parabolic HP (up to second order, green line) is plotted
in the $x-z$-plane along with the expansion to fourth order (blue line) in comparison with the HP according to Eq.(5) shown as the red 
line.
\begin{figure}[!h]
\begin{center}
 \includegraphics[width=0.65\textwidth]{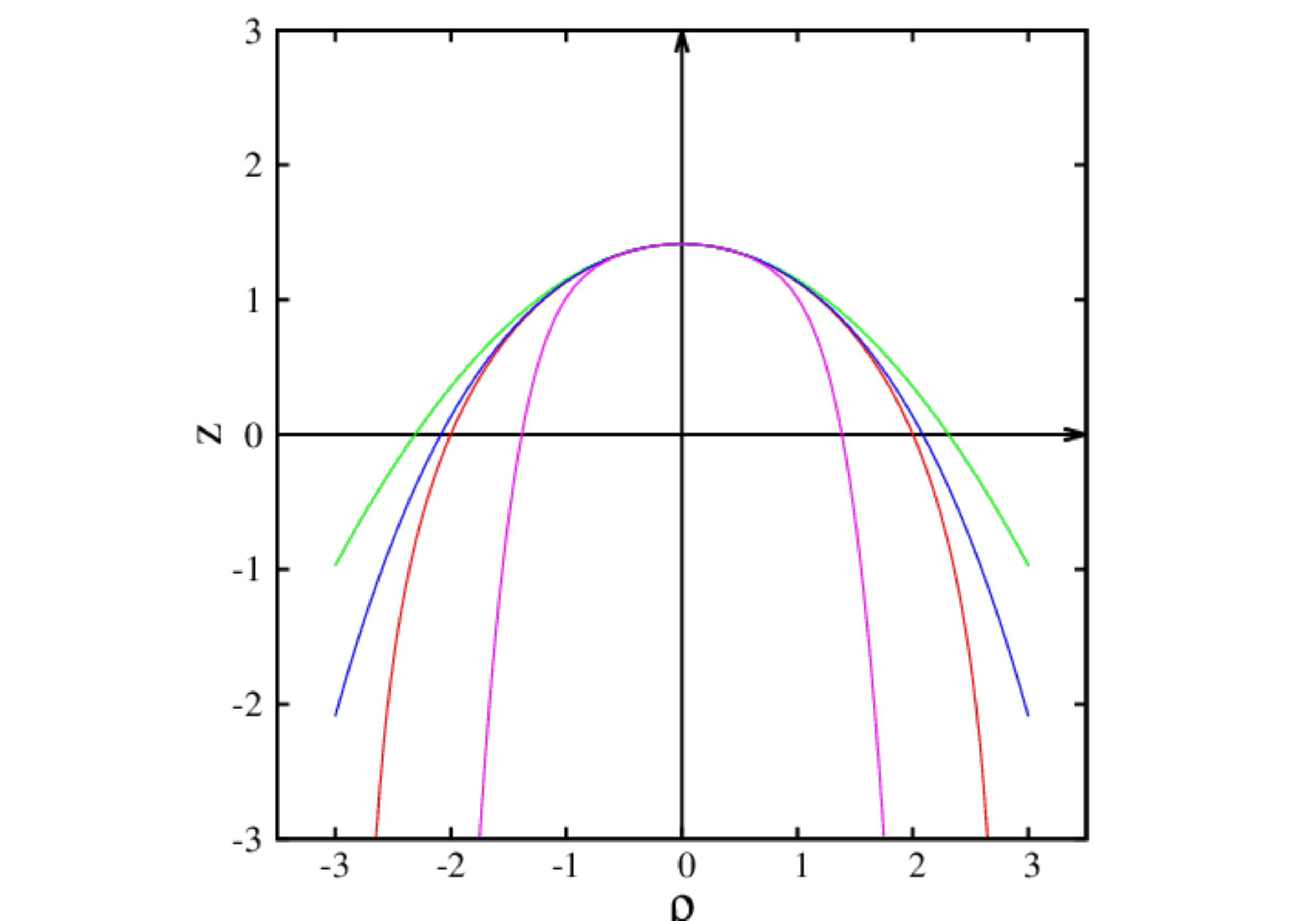}
 \caption{The HP curve defined with Eq.(5) (red line), its Taylor approximations up to second (green line) and fourth (blue line) order 
          and an extremely narrow HP.}
\end{center}
\end{figure}
Evidently, even in lowest order the approximation is already reasonable in the upwind heliosphere, i.e. for positive $z$. By multiplying one
or more Taylor coefficients with factors greater than unity a narrower HP can be obtained, like the example illustrated with the violet line
(obtained by multiplying the coefficient of the sixth order by 100). These HP shapes are not only strongly deviating from the HP function Eq.(5)
and are, thus, strongly inconsistent with the flow field, but they still, like all polynomial surfaces, are characterized by a diverging cross 
section of the heliotail, an undesired feature that the HP defined with Eq.(5) avoids. 
\end{document}